\documentclass[twocolumn,pre,showpacs]{revtex4}

\usepackage{amssymb}

\usepackage{epsfig}
\usepackage[usenames]{color}
\newcounter{counter}

\begin{document}

\title{Poiseuille flow of soft glasses in narrow channels: From quiescence
to steady state}

\author{Pinaki Chaudhuri}
\affiliation{Institut f\"ur Theoretische Physik II, Heinrich-Heine-Universit\"at 
D\"usseldorf, 40225 D\"usseldorf, Germany}
\author {J\"urgen Horbach}
\affiliation{Institut f\"ur Theoretische Physik II, Heinrich-Heine-Universit\"at 
D\"usseldorf, 40225 D\"usseldorf, Germany}

\begin{abstract}
Using numerical simulations, the onset of Poiseuille flow in a confined
soft glass is investigated. Starting from the quiescent state,  steady
flow sets in at a timescale which increases with decrease in applied
forcing.  At this onset timescale,  a rapid transition occurs via the
simultaneous fluidisation of regions having different local stresses. In
the absence of steady flow at long times, creep is observed even in
regions where the local stress is larger than the bulk yielding threshold.
Finally, we show that the timescale to attain steady flow depends strongly
on the history of the initial state.
\end{abstract}

\maketitle

{\it Introduction.} The advent of microfluidic devices has led to 
studying the structural and dynamical properties of soft materials in
narrow confinements \cite{tabelingbook,colinmicro2012}.  Due to a wide range of
applications, understanding their rheological properties is of great interest.
A typical flow pattern in microchannels is Poiseuille flow which is associated
with a {\it spatially inhomogeneous} stress field. 
Particularly interesting in this context are glassy materials, 
which have a characteristic stress threshold for yielding. Their response to 
inhomogeneous stress fields is, however, not well studied. The main focus
has so far been on studying the response to  spatially uniform stress fields 
\cite{thibaut,thomas,pinaki2,fielding2,damien,papenkort,marco}.

A natural question, which has not been addressed so far, is the
microscopic development of Poiseuille flow from the initial quiescent glass
state towards the approach of steady-state flow, especially in the vicinity of
the yielding threshold $\sigma_d$. 
So far, most studies have focused on the steady
state Poiseuille flow \cite{isa,goyon1,goyon2,durian,durian2,jop,nicolas,vincent,pinaki1}, which 
have shown that the narrow confinement of micro-channels results in 
the local steady state rheology  significantly
deviating from the bulk behaviour \cite{goyon1,goyon2, jop,pinaki1}.
In particular, the naive expectation fails that flow will only occur at regions
where the local stress $\sigma_{\rm loc} > \sigma_{\mathrm d}$. Dynamic fluctuations were even observed at
the centre of the channel where $\sigma_{\rm loc} < \sigma_{\mathrm d}$
\cite{jop,nicolas,vincent}. These findings have been associated with non-local
processes that drive the flow in soft amorphous assemblies
\cite{bocquet,kamrin}.

In this Rapid Communication, we study the onset of planar Poiseuille flow in a model colloidal glass
confined between rough walls.  We demonstrate that
the imposed inhomogeneous stress field in combination with strong confinement leads to the 
following observation: there is a sudden and rapid
transition to steady flow, after a waiting time which depends on the applied
external forcing and the history of the sample. The cooperative nature of the
yielding process is expressed by the nearly simultaneous fluidization of
different regions in the sample, although local stresses vary significantly
across the channel (including $\sigma_{\rm loc}<\sigma_{\mathrm d}$).  At small
forcings, spatially resolved dynamical measurements reveal a creep flow regime,
even in regions where $\sigma_{\rm loc}>\sigma_{\mathrm d}$.  
For wider channels, we however, demonstrate that the transient dynamics 
is qualitatively different. Thus, our study
reveals that the interplay between stress inhomogeneities and cooperative
behavior results in a surprising transient flow response of glasses in the
Poiseuille flow geometry.

{\it Simulation details.} We consider the model colloidal system
of a $50:50$ binary Yukawa fluid (for details, see \cite{zausch1,zausch2,winter}).  
Molecular dynamics simulations were done at constant particle number density
$\rho=N/V=0.676\,d_s^{-3}$ in the canonical ensemble for samples
consisting of $N=12800$ particles in a box of volume $V=L_x \times L_y
\times L_z$, with the dimensions $L_x=53.32\,d_s$, $L_y=26.66\,d_s$, and
$L_z=13.33\,d_s$ (here, $d_s$ is the diameter of the smaller particles in
the binary mixture with size ratio 1.2). The
temperature $T$ was controlled via a Lowe thermostat \cite{lowe1,lowe2}.
At a high temperature of $T_0=0.2$, the system was equilibrated using
periodic boundary conditions. Then, $m=40$ independent configurations
sampled at this $T_0$ were instantaneously quenched to $T=0.10$ (below the
mode-coupling critical temperature of $T_c=0.14$ \cite{zausch1,zausch2}).
After each of these $m$ configurations were aged for durations of
$t_{\mathrm{age}}=10^4$, the particles were frozen at $0<y<2\,d_s$
and $L_y-2\,d_s<y<L_y$ to obtain glassy states confined between rough
walls. Thus, the effective width of the channels (i.e.~the wall-to-wall
distance) was $w=22.66\,d_s$.  This width, typical to experimental
micro-channels, is narrow enough to observe significant deviations from
bulk rheological behavior \cite{isa, goyon1}. For comparison of flow
response, we also consider widths of $w/d_s=49.33, 102.66$, using
same number of particles.

Inside the channel, Poiseuille flow is set up by applying a body force,
$F_0\hat{x}$, on each particle \cite{varnikpois}.  For this external
forcing, the local stress in the velocity-gradient plane is given by
$|\sigma(y)|=\rho{F_0}|y-w/2|$ \cite{toddevanspois}. 
Since $\sigma(y)$ is a monotonic function,
all forcings are quantified by the stress at the wall, $\sigma_{\mathrm
w}=\frac{1}{2}\rho{F_0}w$, scaled by the dynamic yield stress
$\sigma_{\mathrm d}$ \cite{epaps}.

\begin{figure}
{\includegraphics[scale=0.45]{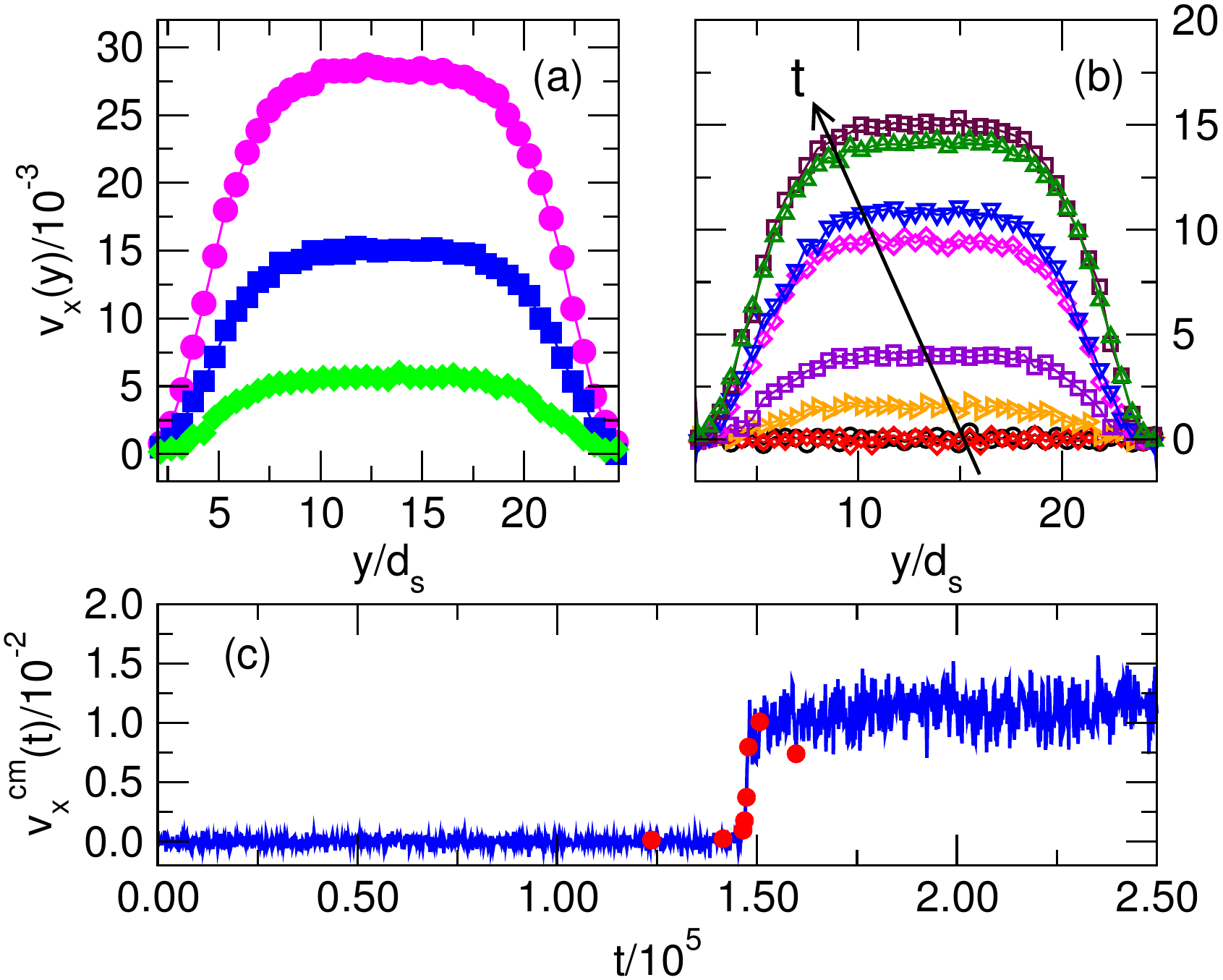}}\\
{\includegraphics[scale=0.7]{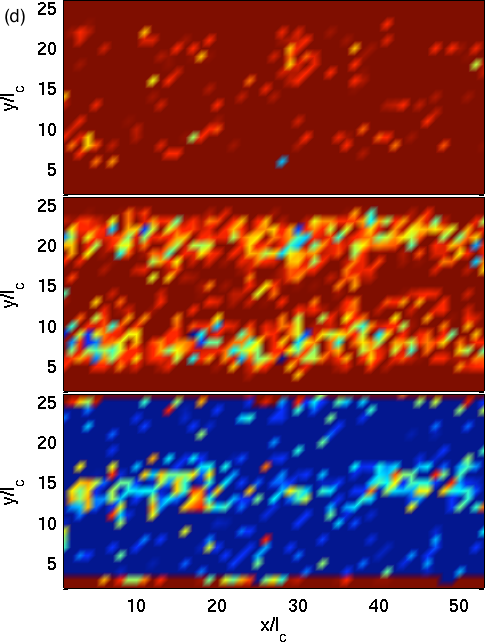}}
{\includegraphics[scale=0.6]{fig2b.pdf}}\\
\caption{(a) Steady state velocity profiles for $\sigma_{\mathrm
w}/\sigma_{\mathrm d}=2.65$ (magenta), 2.37 (blue), 2.09 (green). (b)
Transient evolution of $v_x(y)$ before onset of steady state,
for $\sigma_{\mathrm w}=2.37\sigma_{\mathrm d}$. (c) Variation of the
spatially averaged flow velocity indicating the sudden outbreak of flow
at $\sigma_{\mathrm w}=2.37\sigma_{\mathrm d}$.
(d) Maps of transverse displacements at $t/10^5=1.23,1.46,1.59$.  
The colorbar shows the displacement scales (in units of $d_s$).
}
\label{fig1a}
\end{figure}

{\it Velocity profiles.} Once steady state is reached, 
we obtain the spatial profiles of
the local flow velocities, $\langle{v_{x}(y)}\rangle$, as shown in 
Fig.~\ref{fig1a}(a), with $\langle \cdots \rangle$ denoting a steady state average. 
Unlike the plug-like profiles observed for Poiseuille flow of glassy 
systems in wider channels \cite{isa,varnikpois,papenkort}, 
the velocity profiles, shown here, have a  more rounded central region, 
similar to earlier studies in narrow channels
\cite{goyon1,goyon2,vincent}. 

The time evolution of the velocity profiles, for a single initial
configuration, from the quiescent glass
state at $t=0$ to the steady state is monitored in Fig.~\ref{fig1a}(b) for $\sigma_{\mathrm w}=2.37\sigma_{\mathrm d}$. 
Each of the velocity profiles are averaged over a short time period,
$\Delta{t}=415$, centered around successively increasing times of measurement.
We observe that, at first, no flow is visible, until a point in time when
for regions near the walls, the local shear rates (i.e. $\partial{v_x}/\partial{y}$) become finite
and first signs of a spatially varying $v_{x}(y)$
are seen, albeit small in magnitude. With increasing time, the
magnitude of the local velocities increases until the steady state profile
finally sets in.

Further information about the evolution to steady state is provided by the
centre-of-mass velocity in $x$-direction, $v_{x}^{\rm cm}(t)$, corresponding to
spatial average of the instantaneous velocity profile $v_{x}(y)$. When flow
occurs, $v_{x}^{\rm cm}$ displays a finite steady-state value; see Fig.~\ref{fig1a}(c). 
Thus,   for $\sigma_{\mathrm w}=2.37\sigma_{\mathrm d}$ there is no
flow in the channel for a long time and then suddenly, there is a burst of flow
leading to a quick approach of the steady state.  The velocity profiles shown
in Fig.~\ref{fig1a}(b) are sampled during the narrow time window over which
this transition occurs (as marked with circles in Fig.~\ref{fig1a}(c)). This
indicates that the fluidization occurs in almost all regions inside the channel
within a short time-window.

\begin{figure}
{\includegraphics[scale=0.37,clip=true]{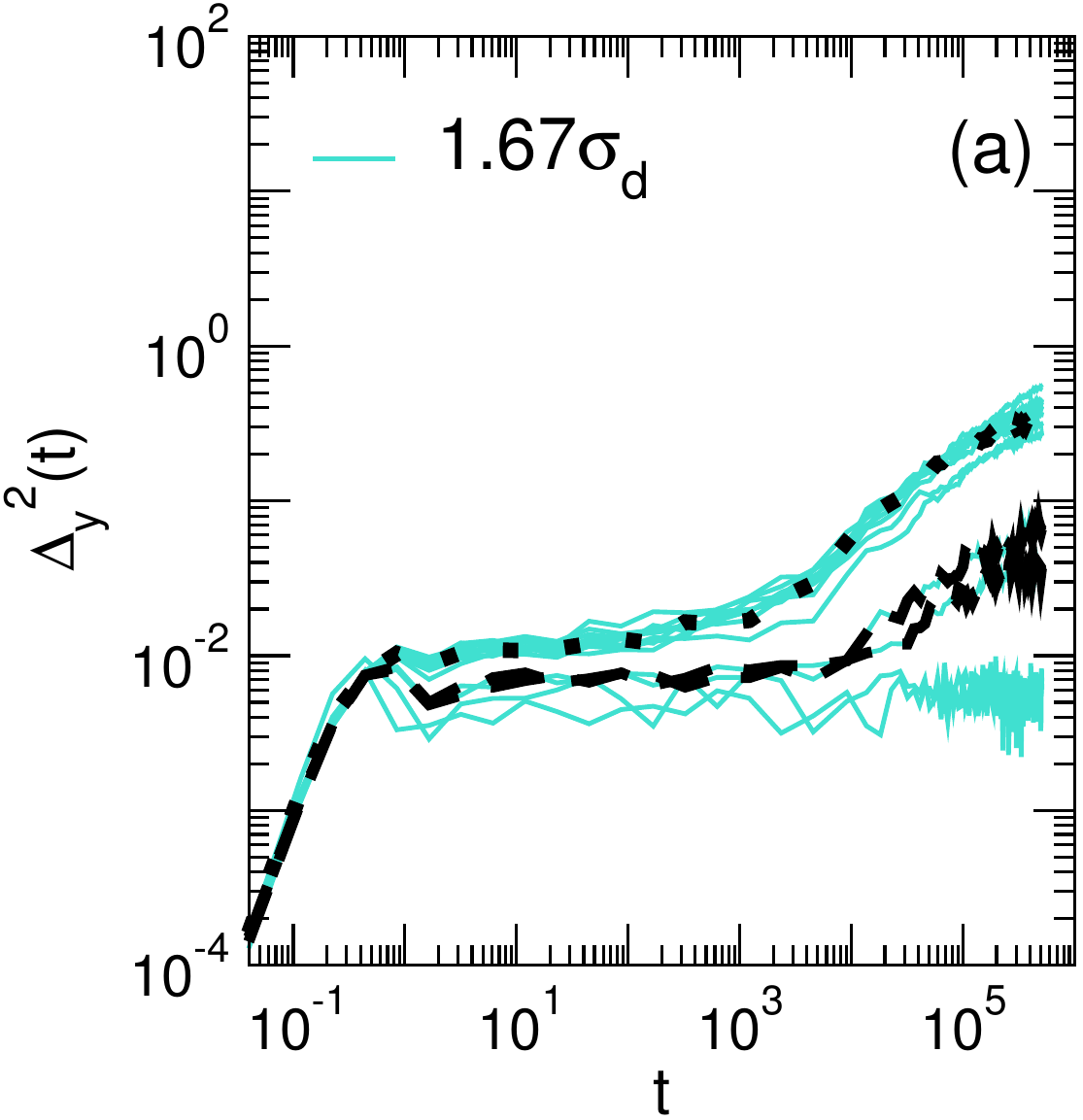}}
{\includegraphics[scale=0.37,clip=true]{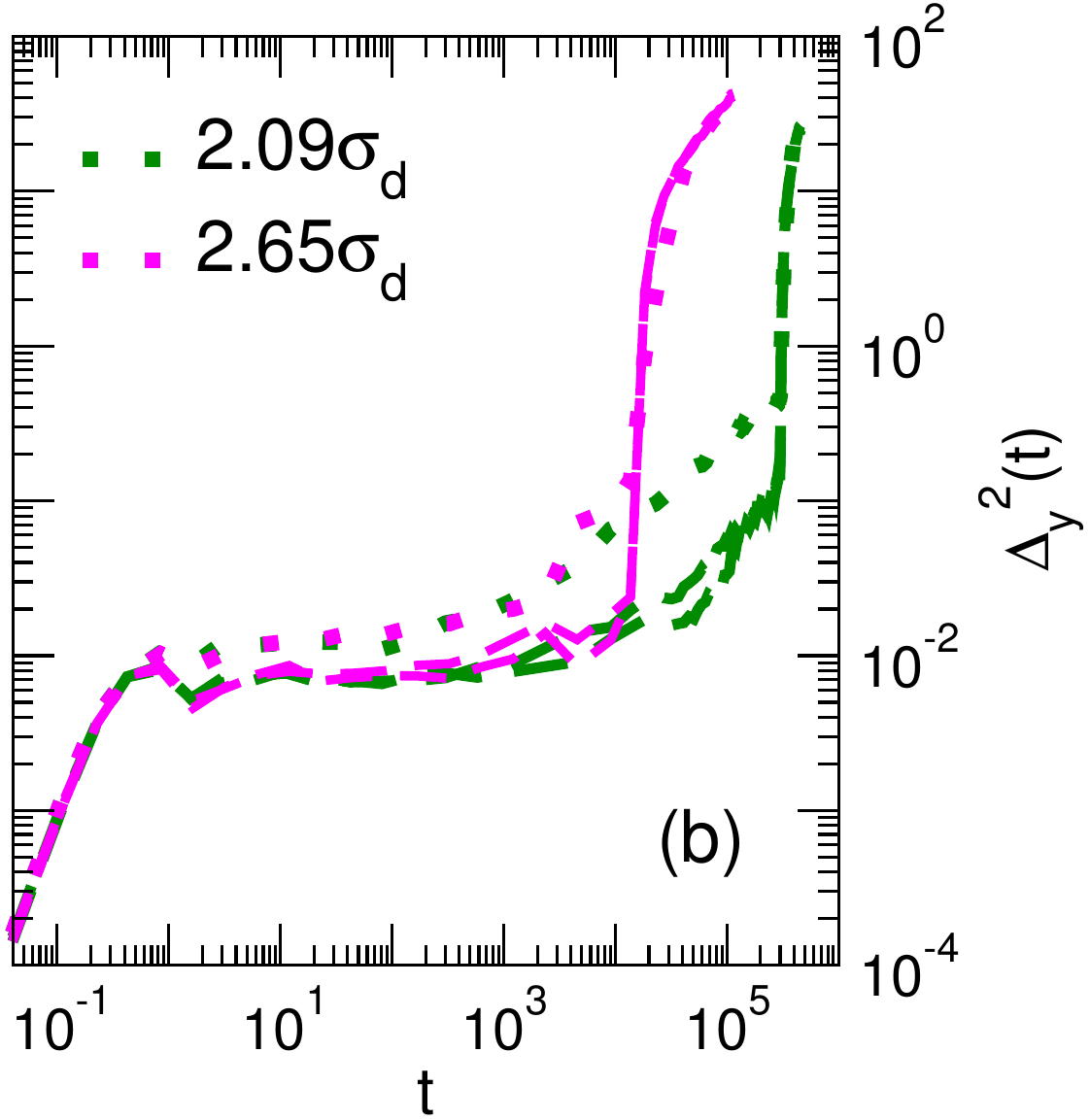}}\\
\caption{
(a) During flow onset, layer-resolved MSD $\Delta_y^2(t)$ for
$\sigma_{\mathrm w}=1.67\sigma_{\mathrm d}$, marking data for
the central layer (dotted), and layers centred $1d_s$ away from the two walls (dashed).
(b) for $\sigma_{\mathrm w}/\sigma_{\mathrm d}=2.09$ and 2.65, MSD for the latter three layers. 
}
\label{fig2a}
\end{figure}

{\it Mobility maps.} In order to observe how the flow develops at a local
scale, we construct maps of local transverse displacements \cite{pinaki2},
which are non-affine motions caused by local structural rearrangements
(see Supp. Mat. \cite{epaps}).

In Fig.~\ref{fig1a}(d), we illustrate how the local displacements evolve
during the time period over which flow suddenly sets in, as marked
in Fig.~\ref{fig1a}(c).  We begin at $t=1.23\times{10^5}$, when the
first signs of a spatially varying velocity profile is emerging [see
Fig.~\ref{fig1a}(b)]. The local dynamics reveals a few spots of enhanced
mobility scattered across space, albeit a few more at points far from
the central plane. At a later time ($t=1.46\times{10^5}$), 
these spots have evolved into regions of increased mobility in the
planes which have higher local stress, albeit at regions shifted from
the walls. At this time, the velocity profile is somewhere in the middle
of its transition toward steady state, see Fig.~\ref{fig1a}(b).
When steady flow has emerged ($t=1.59\times{10^5}$),
the entire channel has been fluidized including the region at the centre
(where  $\sigma_{\rm loc} < \sigma_{\mathrm d}$).

{\it Mean squared displacements.} To further elucidate the motion
of particles, which are initially located at regions having different
local stresses, we monitor the mean squared displacements (MSD) of single
particles. Similar to the construction of the displacement maps, at $t=0$,
we subdivide the $xy$ plane into slabs of thickness $l_c=2.05\,d_s$ and
identify the particles in each such slab. Subsequently, we calculate
the average MSD ($\Delta_y^2(t)$) of the particles originating in each slab for motions
in the transverse direction, i.e. along the direction of stress
gradient (see Supp. Mat. \cite{epaps}). In Fig.~\ref{fig2a}, we show the
MSD data for an initial configuration 
subjected to various external forcings, to
compare how a glassy state responds to increasing $\sigma_{\mathrm w}$.
For clarity, we also mark the data for particles  located in the central
slab and the slabs near the two walls.

For $\sigma_{\mathrm
w}=1.67\sigma_{\mathrm d}$ (Fig.~\ref{fig2a}(a)), the dynamics is
sub-diffusive in all the three layers over the time of observation. This suggests
the existence of a creep flow \cite{coussot, thomas}. Thus, even for
regions where $\sigma_{\rm loc} > \sigma_{\mathrm d}$, i.e.~for $6.7{d_s}
< |y-w/2| < w/2$, we observe creep. Such motion is not seen in athermal
jammed systems, where the material quickly reaches either a completely
stuck or a flowing state \cite{pinaki1, ciamarra}, indicating the significance
of thermal fluctuations to the occurrence of creep.

In Fig.~\ref{fig2a}(b), we show the data for  $\sigma_{\mathrm
w}/\sigma_{\mathrm d}=2.09$ and 2.65, focusing again on the three 
layers mentioned above.  In both cases, there is
the identical initial caging regime, followed by the dramatic burst into
flow at a timescale which decreases with increasing $\sigma_{\mathrm w}$.
Therefore, whenever there is steady flow at
long times, diffusive dynamics is observed for particles originating
from each of these layers, including those where $\sigma_{\rm loc} <
\sigma_{\mathrm d}$ (see Suppl. Matt.~\cite{epaps}). For $\sigma_{\mathrm w}=2.09\sigma_{\mathrm d}$, 
e.g., this corresponds to $0 < |y-w/2| < 5.37{d_s}$, i.e.
nearly half the channel width. Also, for all the layers, the jump in the
dynamics occurs at nearly the same timescale.  
Thus, this complements the scenario of complete fluidisation across the channel as illustrated in
the displacement maps of Fig.~\ref{fig1a}.  The displacement data
conclusively demonstrates the correlated processes underlying the yielding
of such materials. Regions having different local stresses either
fluidize (i.e. attain steady flow) or continue to creep, depending on the imposed stress gradient
which links and constrains the motion across the channel-width.

\begin{figure}[h]
{\includegraphics[scale=0.375]{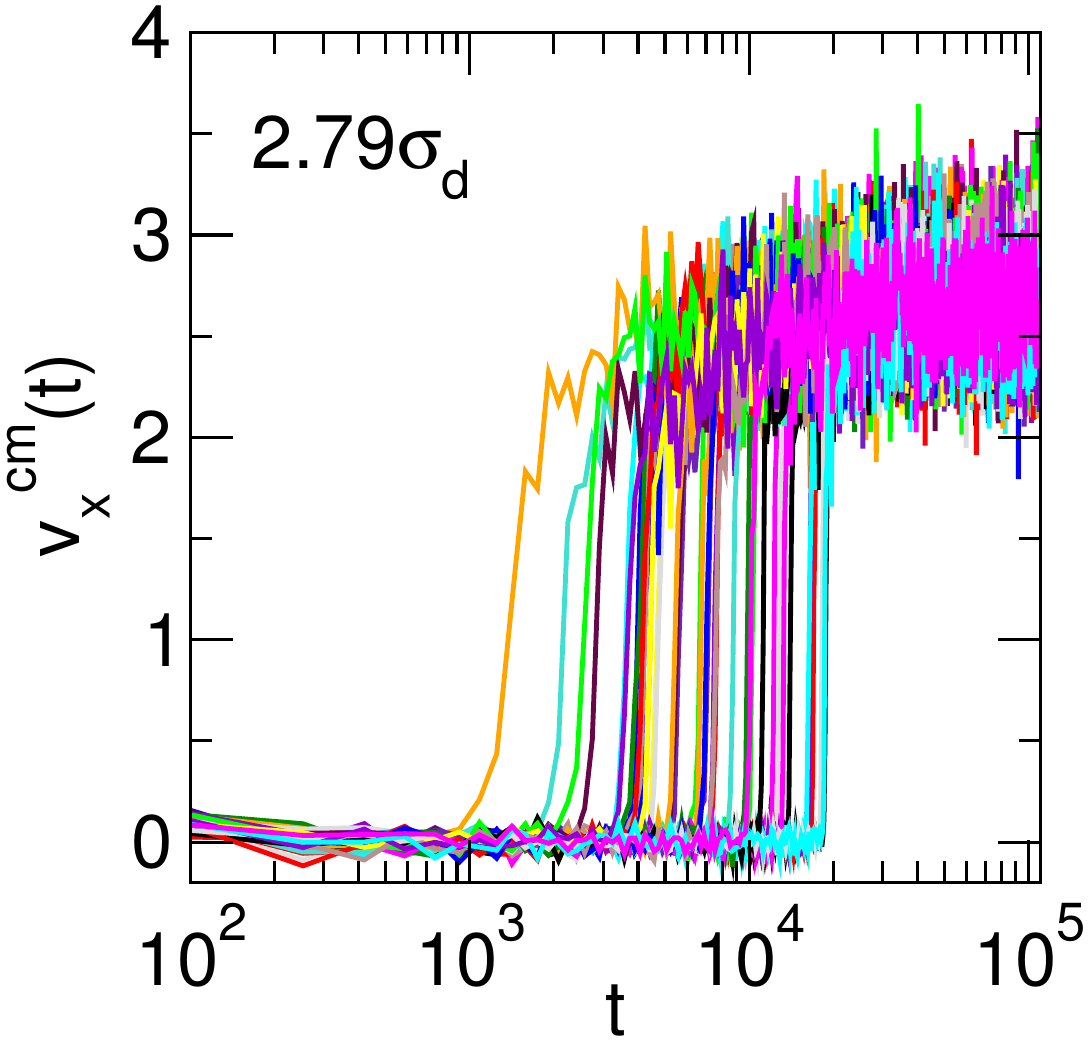}}
{\includegraphics[scale=0.375]{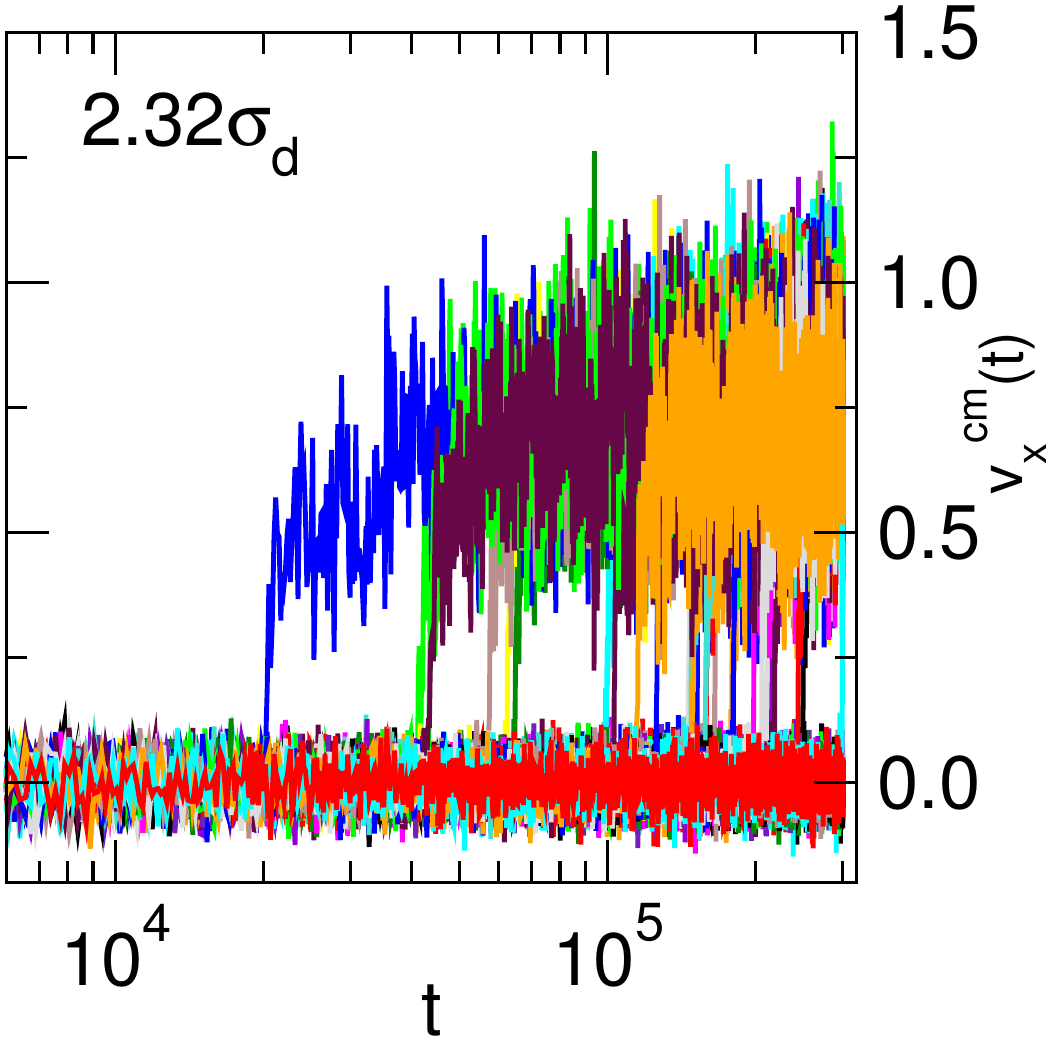}}\\
{\includegraphics[width=3.2in]{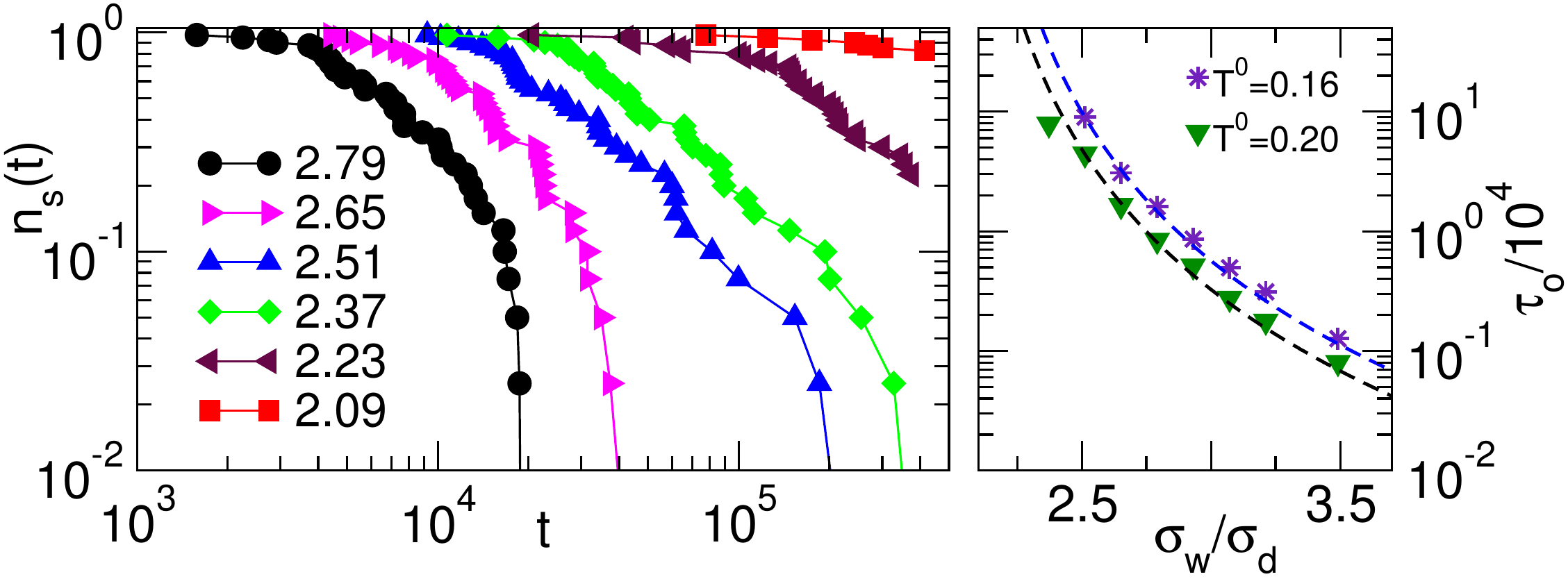}}
\caption{({\it Top}): Flow velocity, $v_x^{cm}(t)$, for $m=40$ initial
states ($t_{\rm age}=10^4$), prepared via quench from $T_0=0.20$, 
at $\sigma_{\mathrm w}/\sigma_{\mathrm d}=2.79$ (left), $2.23$ (right).  
({\it Bottom}): (left) For the same initial states, time evolution
of fraction of non-flowing states $n_s(t)$ for different $\sigma_{\mathrm
w}$. (right) Variation of $\tau_{o}$ with $\sigma_{\mathrm w}$ for initial
states, prepared via quench from $T_0=0.2$ (green triangles) and $0.16$  (purple star). 
The dashed line is a fit with ${A}/{(\sigma_{\rm w}/\sigma_d - x_c)^\beta};
x_c=2.01, 2.05$ for the two respective $T_0$.} 
\label{fig4}
\end{figure}

{\it Time-scales for onset of flow.} Until now, we have studied the
spatio-temporal evolution of the flow
for a single initial configuration. Now, we expand the analysis to the
ensemble of $m=40$ initial configurations. In Fig.~\ref{fig4}(a)-(b),
we show the flow velocity $v_x^{\rm cm}(t)$ for all the $m$ trajectories
within the ensemble at imposed stresses of $\sigma_{\mathrm w}/\sigma_{\rm
d}=2.79$, 2.23. For $\sigma_{\mathrm w}=2.23\sigma_{\rm d}$, in all
cases, steady flow is observed at long times and the onset time-scale
depends on the initial condition. However, for a decreased forcing
of $\sigma_{\mathrm w}=2.23\sigma_{\rm d}$, we observe that not only
does the onset time-scales become larger, but also a certain fraction
of the trajectories undergo the transition to steady flow during the
period of our observation. This is further illustrated for different
$\sigma_{\mathrm w}$ in Fig.~\ref{fig4}(c), where we plot the time
evolution of the fraction of states that have not yielded, $n_s(t)$.
We observe that for $\sigma_{\mathrm w}=2.09\sigma_{\rm d}$, most of
the states remain unyielded resulting in a very slow decay of $n_s(t)$.
Now, $n_s(t)$ is related to the distribution of time-scales $\Pi(t_o)$
for the onset of flow: $n_s(t)=1-\int_{0}^{t}\Pi(t_o) dt_o$. Thus,
with decreasing stress, $\Pi(t_o)$ broadens. 
In Fig.~\ref{fig4}(d), we show (using green triangles) the variation 
of the mean ($\tau_o$) of the distribution of these time-scales with $\sigma_{\mathrm w}$. The
data can be fitted with the function ${A}/{(\sigma_{\rm w}/\sigma_d - x_c)^\beta}$, 
where $A=3136.35, x_c=2.01, \beta=3.83$. We also consider
another ensemble of $m=40$ configurations which were quenched from $T^0=0.16$
and aged for $t_{\rm age}=10^4$. The variation of $\tau_o$ with 
$\sigma_{\mathrm w}/\sigma_{\mathrm d}$ for this set is also shown 
in Fig.~\ref{fig4}(d) (using stars). For each $\sigma_{\mathrm w}/\sigma_{\mathrm d}$,
the typical values are higher than those obtained for the ensemble quenched from $T^0=0.2$.
It is known that quenches from lower supercooled temperatures generate glassy states
with lower energies \cite{kob}. Thus, we observe that longer timescales are necessary
to fluidize these low-lying states. Further, a fit of the timescales provides
a threshold of $x_c=2.05$. Thus, for both ensembles, there is apparent divergence 
at $\sigma_{\mathrm c}\gg\sigma_{\mathrm d}$ 
depending upon the history of quenching. 
This is also different from Couette flow where the divergence occurs at $\sigma_{\mathrm c}\approx{\sigma_{\rm d}}$
\cite{pinaki2, epaps}.

\begin{figure}
{\includegraphics[width=8cm,clip=true]{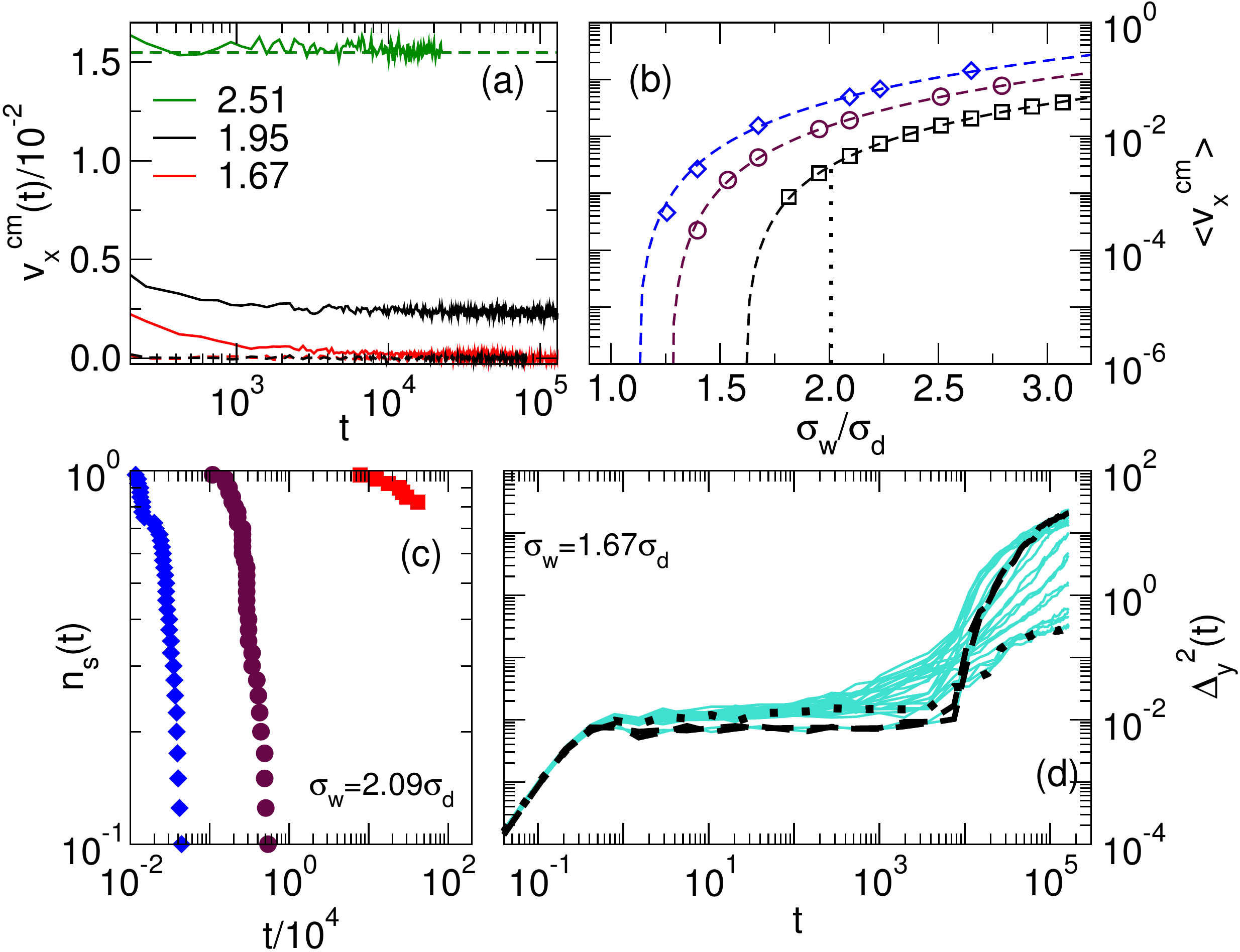}}
\caption{
(a) For $w=22.66d_s$, flow velocities after quench
from  $\sigma_{\mathrm w}=2.79\sigma_{\rm d}$ 
to $\sigma_{\mathrm w}/\sigma_{\rm d}=2.51$, 1.95, 1.67. 
Dashed lines mark steady state averages obtained during
onset of flow.
(b) Variation of steady flow velocity, $\langle{v_x^{cm}}\rangle$, 
with $\sigma_{\mathrm w}/\sigma_{\rm d}$
for  $w/d_s=22.67$ (square), 49.33 (circle), 102.67 (diamond). 
Dashed lines are fits to extract the flow thresholds  $x_t$.
Dotted vertical line marks $x_c$ (see Fig.~\ref{fig4}).
(c) For $\sigma_{\mathrm w}=2.09\sigma_{\mathrm d}$, time evolution
of  $n_s(t)$ for different $w$ (symbols as in (b)). 
(d) For $w/d_s=49.33$, layer-resolved MSDs during the transient regime for
$\sigma_{\mathrm w}=1.67\sigma_{\mathrm d}$, with layers marked similar to Fig.~\ref{fig2a}.} 
\label{fig5}
\end{figure}

{\it Stress quenches.} For glasses, the transient response depends upon 
the preparation of the initial quiescent states
\cite{pinaki2,varnikpois,shi,cates,frahsa13}. Such
dependence on sample history  can
lead to practical advantages for Poiseuille flow.  As noted above, for $\sigma_{\mathrm
w}=2.09\sigma_d$, most states remain non-flowing even at long time-scales.
Thus, at lower $\sigma_{\mathrm w}$,  yielding would be difficult to observe
within experimental time-scales. Now, we demonstrate how steady flow can still
be attained at $\sigma_{\mathrm w} <  2.09\sigma_d$. We take configurations
from the steady state at $\sigma_{\mathrm w}=2.79\sigma_{\rm d}$ and then
suddenly decrease $\sigma_{\mathrm w}$ to a smaller value.  In the left panel
of Fig.~\ref{fig5}, we show the resultant mean flow velocities averaged over
all the $m=40$ independent states. First, we show that for $\sigma_{\mathrm
w}=2.51{\sigma_\mathrm d}$, we recover the same mean flow velocity as observed
during the steady-state obtained from the quiescent state. On the other hand,
if $\sigma_{\mathrm w}=1.67{\sigma_\mathrm d}$ after the quench, no discernible
mean flow is observed at long times, similar to the case when the stresses were
switched on. However, for $\sigma_{\mathrm w}=1.95{\sigma_\mathrm d}$, we do
see a sustained steady flow after quenching, in contrast to the
situation after start-up, and a stable velocity profile is also developed. 
In Fig.\ref{fig5}(b), we plot the variation of steady state average of the 
flow velocity, $\langle{v_x^{\rm cm}}\rangle$, with $\sigma_{\mathrm w}/\sigma_{\mathrm d}$.
The set of data points can be fitted with $B{(\sigma_{\rm w}/\sigma_d - x_t)^\theta}$,
obtaining $x_t=1.61$. Thus, even though this threshold 
for obtaining finite $\langle{v_x^{\rm cm}}\rangle$ is smaller than the threshold 
where $\tau_o$ apparently diverges, it is still larger than $\sigma_d$.

{\it Wider channels.} We also explored how $x_t$ varies with increasing channel width. In Fig.\ref{fig5}(b), 
 $\langle{v_x^{\rm cm}}\rangle$  is shown for $w/d_s=49.33, 102.66$ and similar fits as above
 provides $x_t=1.27,1.11$ respectively, i.e.  the threshold decreases with increasing channel width and 
approaches the bulk limit. A consequence of this decrease
in $x_t$ with increasing $w$ is that the timescale for onset of flow also decreases
for a fixed $\sigma_{\mathrm w}/\sigma_{\rm d}$ (cf. $n_s(t)$ for $\sigma_{\mathrm w}=2.09\sigma_{\mathrm d}$ shown in Fig.\ref{fig5}(c )).
The local dynamics during the onset of flow also seems to change with increasing $w$. 
In  Fig.\ref{fig5}(d), we plot the layer-resolved MSD for $\sigma_{\mathrm w}=1.67\sigma_{\mathrm d}$
in a channel of width $w=49.33d_s$. 
We see that particles in the central layer undergo much smaller displacements compared to those
near the boundaries. Thus, although a weaker stress gradient due to the increasing channel width
allows for flow to occur for this  $\sigma_{\mathrm w}$, the transverse mixing of particle layers
are distinctly different from what is observed for $w=22.66d_s$.

{\it Conclusions and outlook.} Our study shows that
the onset timescales required for steady Poiseuille flow increases with
decrease in applied forcing, and varies strongly from sample to sample
as well as quench history.  For narrow channels,
the onset of the flow occurs simultaneously in all regions of the confined material,
although the local stress varies significantly across the system.  When steady
flow does not set in within observational timescales, the material creeps; this
happens even when local stresses are larger than $\sigma_d$.  This implies that
rather than the individual local stresses, it is the full inhomogeneous stress 
map, characterised by the stress gradient, that determines the development of
steady flow. We also demonstrated that the weakening of the gradient by widening
the channel width influences the flow threshold and also impacts the nature of local
dynamics during yielding.

The rapid onset of flow observed here for the model soft glass 
is reminiscent of granular avalanches
\cite{forterre,amon2013}, albeit in the absence of frictional forces and also
with confining boundaries (rather than free surfaces).  Understanding how such
boundary conditions impact the flow start-up process is necessary
\cite{nicolas2, cloitre2,vincent2}, in particular for even narrower channels 
\cite{travis}. Further work should also explore the yielding process under 
similar stress gradients for not only other soft materials like gels, but 
also for polymeric and metallic glasses; this would have significance for
practical applications.
\begin{acknowledgments}
{\it Acknowledgments.} We thank L. Bocquet, F. Varnik, and T. Voigtmann
for useful discussions. We acknowledge financial support by the Deutsche
Forschungsgemeinschaft (DFG) in the framework of the priority programme
SPP 1594 (grant HO 2231/8-1), and computing time at the NIC J\"ulich.
\end{acknowledgments}
%

%


\vspace{2in}

\begin{center}

{\bf \underline{Supplementary Information}}

\end{center}

\subsection{Measurement of dynamic yield stress $\sigma_{\rm d}$}

The dynamic yield stress was estimated via strain-rate controlled simulations.  
The shear stress ($\sigma$) of the confined glass is measured 
for different imposed shear-rates ($\dot{\gamma}$). The $\sigma$ vs $\dot{\gamma}$
data is fitted with the Herschel-Bulkley function: $\sigma=\sigma_d+B\dot{\gamma}^\alpha$,
which provides us with an estimate of $\sigma_{\rm d}$.

\subsection{Constructing mobility maps}

Before we apply the external forcing (at time $t=0$),
we divide the $xy$ plane of the simulation box into small square cells
(of length $l_c=1.1\,d_s$) and identify the particles in each cell. Next,
after time $t$, we calculate the transverse displacement of each particle
$\Delta{y}_i(t)=|y_i(t)-y_i(0)|$. Then, we construct the maps by
calculating for each cell the local mobility,  
$\mu_{lm}(t)=\langle\Delta{y}_i(t)\rangle_{lm}$, where
$\langle \cdots \rangle_{lm}$ is the average over all the particles in
the cell $\{lm\}$ at $t=0$.

\setcounter{counter}{0}
\renewcommand{\thefigure}{S\arabic{counter}}
\addtocounter {counter}{1}
\begin{figure}[h]
{\includegraphics[scale=0.6,clip=true]{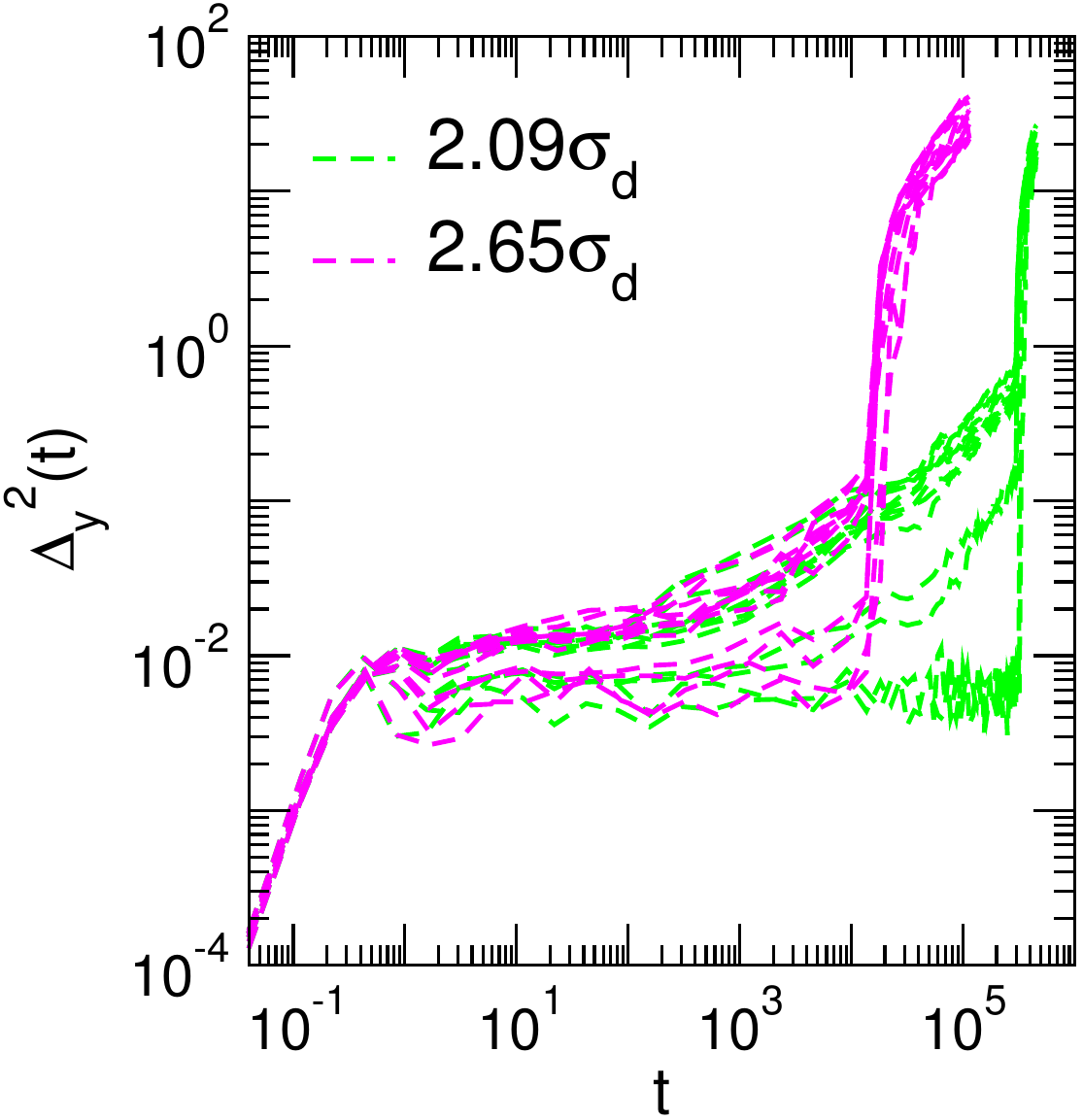}}
\caption{During onset of flow, layer-resolved MSD $\Delta_y^2(t)$ for
$\sigma_{\mathrm w}=2.09$ and 2.65.}
\end{figure}

\subsection{Mean squared displacements: spatially resolved}

At $t=0$, we subdivide the $xy$ plane into slabs which are parallel to the two
confining walls. Each such slab has a thickness of
$l_c=2.05\,d_s$. We identify the particles in each such slab at $t=0$. Subsequently,
we calculate the average MSD of the particles originating in each slab:
$\Delta_y^2(m,t)=\langle[y_i(m,t)-y_i(m,0)]^2\rangle_{n_m}$, where
$\Delta_y^2$ is the MSD calculated for motions in the direction of the stress gradient and the averaging
$\langle \cdots \rangle_{n_m}$ is done for the particles populating the
$m^{th}$ slab at $t=0$. In Fig-S1, we show the transient MSD data, calculated in this manner, for
$\sigma_{\mathrm w}=2.09$ and 2.65, for channel width of $w/d_s=22.66$.

\subsection{Onset timescales : Couette and Poiseuille flow}

For the same ensemble of initial states, quenched from $T_0=0.2$ and
confined within a channel width of $w/d_s=22.66$, we compare
the timescale for onset of flow for two different stress fields, viz.
Couette (spatially uniform) and Poiseuille (spatially non-uniform).
In Fig-S2, we show the onset timescales, $\tau_{\rm onset}$, 
for different applied forcings in the two cases. The plot shows that
the thresholds for apparent divergence of $\tau_{\rm onset}$ are different,
with the Poiseuille flow corresponding to a larger value.

\addtocounter {counter}{1}
\begin{figure}[h]
{\includegraphics[scale=0.7,clip=true]{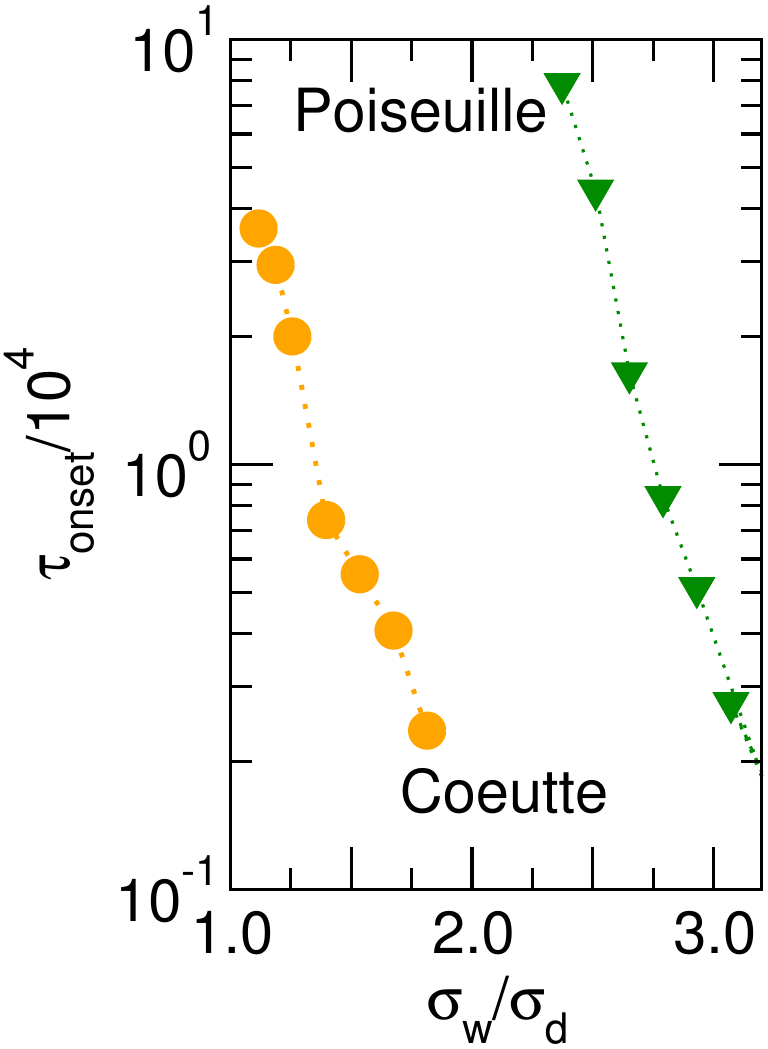}}
\caption{Onset timescales for Couette and Poiseuille flow, for different
appiled forcings $\sigma_{\rm w}$ (in units of $\sigma_{\rm d}$.}
\end{figure}

\end{document}